\documentclass[sn-mathphys]{sn-jnl}

\jyear{2021}%

\theoremstyle{thmstyleone}%
\theoremstyle{thmstyletwo}%

\theoremstyle{thmstylethree}%
\usepackage{color}
\begin{document}

\title[~~]{Generalization of the Kuramoto model to  the Winfree model by a symmetry breaking coupling}


\author[1]{\fnm{M} \sur{Manoranjani}}

\author[2]{\fnm{ Shamik} \sur{Gupta}}

\author[3]{\fnm{D. V.} \sur{ Senthilkumar}}

\author*[1]{\fnm{V. K.} \sur{Chandrasekar}}\email{chandru25nld@gmail.com}

\affil[1]{\orgdiv{Department of Physics}, \orgname{Centre for Nonlinear Science and Engineering, School of Electrical and Electronics Engineering, SASTRA Deemed University}, \orgaddress{ \city{Thanjavur}, \postcode{613 401}, \country{India}}}

\affil[2]{\orgdiv{Department of Theoretical Physics}, \orgname{Tata Institute of Fundamental Research}, \orgaddress{ \city{Mumbai}, \postcode{400005}, \country{India}}}

\affil[3]{\orgdiv{Departmentof Physics}, \orgname{Indian Institute of Science Education and Research}, \orgaddress{ \city{Thiruvananthapuram}, \postcode{695016}, \country{India}}}


\abstract{	We construct a nontrivial generalization of the paradigmatic Kuramoto model by using an additional coupling term that explicitly breaks its rotational symmetry
	resulting in a variant of the Winfree Model.  Consequently,  we observe the characteristic features of the phase diagrams of both the Kuramoto model and the Winfree model
	depending on the degree of the symmetry breaking coupling strength for unimodal frequency distribution.  The phase diagrams of both the Kuramoto and the Winfree models
	resemble each other for symmetric bimodal frequency distribution for a range of the symmetry breaking coupling strength except for region shift and difference in the
	degree of spread of the macroscopic dynamical states and bistable regions.  The dynamical transitions in the bistable states are characterized by an abrupt (first-order)
	transition in both the forward and reverse traces.  For asymmetric bimodal frequency distribution, the onset of bistable regions depends on the degree of  the asymmetry.
	Large degree of  the symmetry breaking coupling strength promotes the synchronized stationary state, while  a large degree of heterogeneity, proportional to the separation 
	between the two central frequencies,  facilitates the spread of the incoherent and standing wave states  in the phase diagram for a low strength of
	the  symmetry breaking coupling.  We  deduce the low-dimensional  equations of motion for the complex order parameters
	using the  Ott-Antonsen ansatz   for both unimodal and bimodal frequency distributions.  
	We also deduce the Hopf, pitchfork,  and saddle-node bifurcation curves  from the  evolution equations for the complex order parameters
	mediating the dynamical transitions.  Simulation results of the original discrete set of  
	equations of the generalized Kuramoto model agree well with the analytical bifurcation curves. }




\maketitle
\section{Introduction}
\label{sec:intro}

Symmetry (translational or rotational) prevailing in the coupled dynamical networks  due to the coupling geometry manifests  in a
wide variety of natural systems and in their intriguing macroscopic dynamical states~\cite{tstp2017}.  Nevertheless, symmetry breaking 
couplings are shown to be a source of a plethora of collective dynamical behavior that are inherent to it
and are  mostly inaccessible with the symmetry preserving couplings.  In particular, networks of the paradigmatic Stuart-Landau oscillators
with symmetry breaking coupling have been employed to unravel  several collective dynamical states that mimic a variety
of collective patterns observed in nature and technology.  
For instance,  symmetry breaking coupling facilitates the transition from the homogeneous to an inhomogeneous steady states~\cite{akev2013},
symmetry breaking interaction has been identified as an essential 
feature for the genesis of partially coherent inhomogeneous spatial patterns,  namely chimera death state~\cite{zamk2014,tb2015,kpvkc2015}.
Multicluster oscillation death  states have been observed in  nonlocally coupled Stuart-Landau oscillators with symmetry breaking coupling~\cite{ismk2015}.
Further, the interplay of  the nonisochronicity parameter and the symmetry breaking coupling   is found to facilitate the onset of 
different variants of chimera death state such as  multichimera death state and periodic chimera death states in 
nonlocally coupled Stuart-Landau oscillators~\cite{kpvkc2016}.
The effect of the symmetry breaking coupling has also been investigated on the phenomenon of reviving oscillations~\cite{wei2017}.
Recently, the effect of the symmetry breaking mean-field coupling on the phenomenon of the aging transition has also been investigated
Conjugate couplings, a symmetry breaking coupling, have also been  widely employed in the literature~\cite{rkrr2007,amit2012,wei2022}.
Note that the  pointed out reports are only a tip of an ice-berg and   not an exhaustive  list of studies that employed  symmetry breaking coupling
using the network of the Stuart-Landau oscillators.

Despite the substantial investigations on the effect of the symmetry breaking coupling in networks of Stuart-Landau oscillators, there is a lacunae in
understanding the nontrivial role of the symmetry breaking coupling in the phase only models, which indeed allows for  exact analytical treatment of
the macroscopic dynamical states in most cases.  In particular, phase models such as  Winfree  and Kuramoto models, and their variants have been
extensively employed in the literature to investigate the emergence of various intriguing collective dynamical states.  Interaction among the phase oscillators in 
the Winfree model  is modeled by  a phase-dependent pulse  function and a sensitive function. The former characterizes   the mean-field, while the latter characterizes  
the response of the individual  oscillators to the mean-field~\cite{w:1,w:2}. Winfree model is one of its kind representing a class of pulse-coupled
biological oscillators such as flashing of fire files~\cite{In2}, applauding audience~\cite{In7} and many more.
Interaction among the phase oscillators in  the Kuramoto model is modeled by  the sine of difference between the phases of the oscillator and  has been widely
employed to investigate the emergence of spontaneous synchronization  in a wide variety of biological, chemical, mechanical and physical, systems~\cite{Kuramoto:1984,Acebron:2005,In1}.  Examples include  cardiac pacemaker~\cite{In3}, 
Josephson junction arrays~\cite{In6}, and  power-grids~\cite{In8}.   

A recent study has generalized the Kuramoto model by  including an additional interaction term that breaks the rotational symmetry of the dynamics explicitly 
and unveiled  a  rich phase diagram  with stationary  and standing wave phases due to the symmetry breaking interaction~\cite{In10}. 
Specifically, the authors have considered unimodal frequency distributions
and revealed the emergence of a stationary state, characterized by time independent amplitude and phase of the complex  Kuramoto order parameter,  facilitated by
the symmetry breaking interaction, which is otherwise absent in the original Kuramoto model that allows for the rotational symmetry of the dynamics. 
Interesting, in this work, we elucidate that the Kuramoto model can be  translated into  the Winfree model by  the introduction of the additional
symmetry breaking coupling and  consequently, one can obtain the phase diagrams of both these models simply by tuning the symmetry breaking parameter $q$,
thereby bridging the dynamics of both the models.  Note that the macroscopic dynamical states of the pulse coupled biological oscillators with
different sensitive functions, characterizing the phase-response-curves of  biological oscillators, are peculiar to  the Winfree model and its generalizations,
which are  far from reach for the Kuramoto model and its variants. In particular, we consider both the unimodal and bimodal frequency distributions
to explore the phase diagrams for various values of the symmetry breaking parameter $q$.  On the one hand, we observe the typical phase diagram of the Kuramoto model
characterized only by  incoherent and standing wave states in the absence of the  symmetry breaking interaction  for the unimodal  frequency distribution. 
On the other hand, we observe the phase diagram with incoherent state, standing wave pattern along with the synchronized stationary state  and bistabilities among them,
a typical nature of the Winfree model, for $q=1$.  For an intermediate and  increasing  value of $q\in(0,1)$, one can find the onset of 
the stationary state and eventually the emergence of
bistability among these states in the phase diagram, and enlargement of the bistable regions  resulting in the phase diagram of the Winfree model. 

All  three states are also observed in both Kuramoto and Winfree models for symmetric bimodal frequency distributions along with 
the region of bistability.  The degree of the spread of
the different macroscopic dynamical states  depends on the strength of the symmetry breaking parameter $q$.  Interestingly, for asymmetric
bimodal frequency distributions,  increase in the degree of asymmetry of the frequency distributions favors the onset of bistable regions even for a rather low values of $q$,
which otherwise cannot be observed with the symmetric  bimodal and unimodal frequency distributions.  We arrive at the phase diagrams by numerical simulation
of the original equations of motion.  We  deduce the reduced low-dimensional evolution equations of motion  for the order parameter using the 
Ott-Antonsen ansatz for both unimodal and bimodal frequency distributions.  We also deduce the Hopf, pitchfork and saddle-node bifurcation curves
from the governing equations of motion for the order parameters, which mediates the dynamical transitions in the phase diagrams.  Homoclinic bifurcation
curve is obtained from the XPPAUT software.

The plan of the paper is as follows. In Sec. II, we generalize the Kuramoto model by introducing a symmetry breaking coupling and elucidate that the latter bridges the Kuramoto model
and the Winfree model.  We deduce the reduced low-dimensional evolution equations for the complex order parameters corresponding to the discrete set of generalized
Kuramoto model using the Ott-Antonsen ansatz  for both unimodal and bimodal frequency distributions in Sec. III.  We also deduce Hopf, pitchfork and saddle-node bifurcation curves
from the evolution equations for the complex order parameters  in Sec. III, mediating the  dynamical transitions among the incoherent, standing wave and synchronized stationary states.
In Sec. IV, we  discuss the observed  dynamical states and their transitions  in the various phase diagrams. Finally, in Sec. VI, we summarize the  results.

\section{Model}
We consider a nontrivial generalization of the Kuramoto  model
by including an interaction term that explicitly breaks the rotational symmetry of the dynamics~\cite{In10}.
The phase $\theta_i$ is governed by the set of N ordinary differential equations (ODEs),
\begin{align}
	\dot{\theta_i}&=\omega_i+\frac{\varepsilon}{N}\sum_{j=1}^{N}\big[\sin(\theta_j-\theta_i)+q\sin(\theta_j+\theta_i)\big],
	\label{eq:km2}
\end{align}
for $i=1, \ldots, N$, where $N\gg 1$. Here $\theta_i(t)$ is the phase of the $i$th oscillator at time $t$, $\varepsilon\ge 0$  is the coupling strength, and $q$ is the strength of the symmetry 
breaking coupling. Note that  Eq.~(\ref{eq:km2})  reduces to the Kuramoto model  by setting $q = 0$ and on identifying $\varepsilon$ with the parameter $K>0$. 
Equation~(\ref{eq:km2}) can also  be viewed as a variant of the celebrated Winfree model~\cite{w1,w2,w3,w4} when $q=1$. The Winfree model takes the form
\begin{align}
	\dot{\theta_i}=\omega_i+Q(\theta_i)\sum_{j=1}^{N}P(\theta_j),
	\label{eq:wf}
\end{align} 
where $P(\theta_j)$ is the phase dependent pulse function and the functional form of the response function $Q(\theta)$ characterizes the  phase-response curves
of certain biological oscillators. From  Eq.~(\ref{eq:km2}), it easy to recognize that $Q(\theta)={-}2\sin(\theta)$ and $P(\theta)=\cos(\theta)$.
It is also evident that the symmetry breaking parameter `q'   bridges the Kuramoto and the Winfree models.
Equation~(\ref{eq:km2}) corresponds to the  Kuramoto model  when $q=0$ and it corresponds to a variant  of the Winfree model when $q=1$, as in Eq.~(\ref{eq:wf}).
We consider the frequencies of the phase-oscillators are distributed  both by  the unimodal Lorentzian distribution given as
\begin{align}
	g(\omega)&=\frac{\gamma}{\pi((\omega-\omega_0)^2+\gamma^2)};~~\gamma >0,
	\label{eq:lor}
\end{align}
and  bimodal Lorentzian distribution represented as
\begin{align}
	g(\omega)&=\frac{1}{\pi}\left[\frac{\gamma_1}{((\omega-\omega_0)^2+\gamma_1^2)}+\frac{\gamma_2}{((\omega+\omega_0)^2+\gamma_2^2)}\right];~~~~~~~~\gamma_1, \gamma_2 >0.
	\label{eq:bil}
\end{align}
Here $\gamma$, $\gamma_1$ and $\gamma_2$  are the width parameter (half width at half maximum)
of the Lorentzian and $\pm\omega_0$ are their central frequencies.  Note that
$\omega_0$  corresponds to the degree of detuning in the system, which is 
proportional to the separation between the two central frequencies.
Note that the bimodal distribution $g(\omega_0)$ is symmetric about zero when  $\gamma_1=\gamma_2$. 
It is also to be noted that $g(\omega_0)$  in Eq.~(\ref{eq:bil}) is bimodal if and only if the  separation between their central
frequencies are sufficiently greater than their widths.  To be precise, it is required that $\omega_0 > \gamma_{1,2}/\sqrt{3}$ for the distribution to be a bimodal, 
otherwise the classical results of the unimodal distribution holds good. 

Heterogeneity in the frequency distribution plays a crucial role in the manifestation of a plethora of collective dynamics 
in a vast variety of natural systems. In particular, coexisting co-rotating and counter-rotating systems characterized by positive and negative frequencies, respectively, 
are wide spread in nature.  For instance, counter-rotating spirals are observed in protoplasm of the Physarum plasmodium~\cite{ref1}, counter-rotating vortices are inevitable in the atmosphere and ocean~\cite{ref2,ref3, ref4}, in magnetohydrodynamics of plasma flow~\cite{ref8},  Bose-Einstein condensates~\cite{ref9,ref10}, and in other physical systems~\cite{ref5,ref6,ref7}.
  Very recently, the counter-rotating frequency induced dynamical effects were also reported in the coupled Stuart-Landau oscillator with symmetry
preserving as well as symmetry breaking couplings~\cite{ref11}. The coexistence of co-rotating and counter-rotating oscillators was initially identified
by Tabor~\cite{ref12}, which is followed by a series of work  employing co-rotating and counter-rotating oscillators.  
All these physical systems strongly suggest that counter-rotating time-evolving dynamical
systems indeed  exist in nature and  play a pertinent  role in the manifestation of their intriguing collective dynamics.

In the following, we will deduce the low-dimensional evolution equations for the 
complex macroscopic order parameters corresponding to both the unimodal and bimodal frequency distributions using the Ott-Antonsen (OA) ansatz~\cite{Ott:2008,Ott:2009}. 
Subsequently, we also deduce the various bifurcation curves facilitating the dynamical transitions  among the observed dynamical states in the phase diagrams.

\section{Low-dimensional evolution equations for the macroscopic order parameters}
We now provide an analysis of the dynamics~(\ref{eq:km2}), in the limit
$N \to \infty$, by invoking the Ott-Antonsen ansatz. In this limit, the dynamics of the discrete set of equations (\ref{eq:km2}) can be
captured by the probability distribution function $f(\theta,\omega,t)$, defined such that $f(\theta,\omega,t){\rm d}\theta$
gives the probability of oscillators with phase in the range
$[\theta,\theta+{\rm d}\theta]$ at time $t$. The distribution is
$2\pi$-periodic in $\theta$ and obeys the normalization 
\begin{equation}
	\int_0^{2\pi} {\rm d}\theta~f(\theta,\omega,t)=g(\omega)~\forall~\omega.
	\label{eq:norm}
\end{equation}
Since the dynamics (\ref{eq:km2}) conserves the number of 
oscillators with a given $\omega$, the time evolution of $f$ follows the
continuity equation
\begin{equation}
	\frac{\partial f}{\partial t}+\frac{\partial(fv) }{\partial
		\theta}=0,
	\label{eq:continuity-equation}
\end{equation}
where $v(\theta,\omega,t)$ is the angular velocity of the oscillators. From Eq. (\ref{eq:km2}), we have,
\begin{equation}
	v(\theta,\omega,t)=\omega+\frac{\varepsilon}{2i}[(ze^{-i\theta}-z^\star e^{i\theta})+q(ze^{i\theta}-z^\star e^{-i\theta})],
\end{equation}
where $z^\star$ denotes  the complex conjugate of the macroscopic order parameter defined as
\begin{equation}
	z=\int_{-\infty}^{\infty} g(\omega) \int_0^{2\pi}   f(\theta, \omega, t)e^{i\theta}d\theta d\omega.
	\label{eq:mo}	
\end{equation}
Now, $f(\theta, \omega, t)$ can be expanded in terms of Fourier series of the form
\begin{equation}
	f(\theta,\omega,t)=\frac{g(\omega)}{2\pi}\left[1+\sum_{n=1}^\infty
	\left(\alpha_n(\omega,t) e^{i n\theta}+{\rm c.c.}\right)\right],
	\label{eq:f-Fourier}
\end{equation}
where, $\alpha_n(\omega,t)$ is the $n$th Fourier
coefficient, while c.c. denotes  complex conjugation of the preceding sum
within the brackets. The normalization condition in
(\ref{eq:norm}) is satisfied by the presence of the prefactor of $g(\omega)$ in (\ref{eq:f-Fourier}).
The Ott-Antonsen ansatz consists in assuming~\cite{Ott:2008,Ott:2009} 
\begin{equation}
	\alpha_n(\omega,t)=\left[\alpha(\omega,t)\right]^n.
	\label{eq:OA}
\end{equation}
Now, it is straightforward to  obtain
\begin{equation}
	\frac{\partial\alpha}{\partial t}+i\omega\alpha+\frac{\varepsilon_1}{2}\left[(z\alpha^2-z^\star)+q(z-z^\star\alpha^2)\right],
	\label{eq:12}
\end{equation}
where,
\begin{equation}
	z^\star=\int_{-\infty}^{\infty}\alpha(t,\omega)g(\omega)d\omega.
	\label{eq:13}	
\end{equation}
\subsection{Unimodal Distribution}
\label{sec:ud}
Substituting  the partial fraction expansion of the unimodal frequency distribution $g(\omega)$ (\ref{eq:lor}) in Eq.~(\ref{eq:13}) and evaluating the
integral using  an appropriate contour integral, one can obtain the order parameter as
\begin{equation}
	z(t)=a^\star(\omega_0-i\gamma,t).
	\label{eqod}
\end{equation}
From (\ref{eq:12}) and (\ref{eqod}), one can obtain the evolution equation for the complex order parameter as
\begin{align}
	\frac{\partial z}{\partial t}-{
		i}(\omega_0+i\gamma)z+\frac{\varepsilon_1}{2}\bigg[\big[|z|^2z-z\big]+{q}\big[z^\star -z^3\big]\bigg]=0.
	\label{eq:z-dynamics}
\end{align}
The above evolution equation for  the complex order parameter $z(t)=r(t)e^{i\psi(t)}$ can be expressed in terms of the evolution equations in $r$ and
$\psi$ as
\begin{subequations}
	\begin{eqnarray}
		&\frac{{\rm d}r}{{\rm d}t}&=-\gamma
		r-\frac{r\varepsilon}{2}(r^2-1)(1-q\cos(2\psi)),\\ 
		&\frac{{\rm d}\psi}{{\rm
				d}t}&=\omega_0+\frac{{\varepsilon}q}{2}(r^2+1)\sin(2\psi)).
	\end{eqnarray}
	\label{eq:r-dynamics}
\end{subequations}
The above equations govern the reduced  low-dimensional order parameter
dynamics, which actually corresponds to the dynamics of the original discrete set of equations~(\ref{eq:km2}) in the limit $N
\to \infty$  for the unimodal Lorentzian distribution function $g(\omega)$~(\ref{eq:lor}). Now, we discuss the various  asymptotic macroscopic dynamical states
admitted by Eq.~(\ref{eq:r-dynamics}). 
\subsubsection{Incoherent (IC) state:}
\label{subsec:iss}

The incoherent (IC) state is characterized by time independent $z$ satisfying
$z=z^\star=0$ (thus representing a stationary state of the
dynamics~(\ref{eq:r-dynamics})); correspondingly, one has $r=0$.
The linear stability of such a state is determined by linearizing
Eq.~(\ref{eq:z-dynamics}) around $z=0$.  By  representing   $z=u$ with $|u|\ll$1, we obtain 
\begin{equation}
	\frac{\partial u}{\partial t}+(\gamma-i \omega_0)u-\frac{\varepsilon}{2}\big[(u)-q(u^{\star})\big]=0.
	\label{eq:u-dynamics}
\end{equation}
Decomposing  $u=u_x + i u_y$ yields
\begin{equation}
	\frac{\partial }{\partial t}
	\begin{bmatrix}
		u_x \\
		u_y
	\end{bmatrix}
	=M
	\begin{bmatrix}
		u_x \\
		u_y
	\end{bmatrix};\\
\end{equation}
\begin{equation}
	~~M \equiv \begin{bmatrix}
		-\gamma+\frac{\varepsilon}{2}\big[1-q\big] & -\omega_0 \\\\
		~\omega_0& -\gamma+\frac{\varepsilon}{2}\big[1+q\big] \\
	\end{bmatrix}.
	\label{eq:M-matrix} \nonumber
\end{equation}
The matrix $M$ has the characteristic eigenvalues
\begin{equation}
	\lambda_{1,2}=\frac{-2\gamma+\varepsilon \pm \sqrt{\Delta}}{2},
	\label{eq:M-eigenvalues}
\end{equation}
with
$\Delta=(\varepsilon^2q^2-4\omega_0^2)$. Note that we have $\lambda_1 > \lambda_2$. The
stability threshold for the incoherent state is then obtained by analysing $\lambda_1$
as a function of $\varepsilon$ and $q$, and seeking 
the particular value of $\varepsilon$ at which $\lambda_1$
vanishes for a given  $q$. The stability threshold can be  obtained  as
\begin{align}
	&\varepsilon_{HB}=2 \gamma ,~~~~~~~~~~~~~~~~   ~~~~~~~~~~~~ \;\;\;
	\mbox{for}\;\; \Delta \le 0, \label{hb}\\
	&\varepsilon_{PF}=2 \sqrt{\frac{\gamma^2+\omega_0^2}{1+q^2}} ~~~~~~~~~~~~~~~~~~~~ \mbox{for}
	\;\; \Delta>0. \label{eq:ISS} 
\end{align}
\subsubsection{Synchronized stationary state (SSS):}
\label{subsec:sss}
Now, we explore the
possibility of existence of the synchronized stationary state.
Requiring  that $r$ and $\psi$ have time-independent non-zero
values  in this case and hence equating the left hand side of equations  (\ref{eq:r-dynamics})
to zero, we obtain the two coupled equations for the synchronized stationary state as
\begin{subequations}
	\begin{eqnarray}
		&\frac{\varepsilon q}{2}\cos(2\psi)&=\frac{\gamma}{(r^2-1)}+\frac{\varepsilon}{2}, \\
		&\frac{\varepsilon q}{2}\sin(2\psi)&=-\frac{\omega_0}{(r^2+1)}.  
		\label{eq:SSS-dynamics}
	\end{eqnarray}
\end{subequations}
With some algebra, one can obtained the following expressions for the stationary $r$ and
$\psi$:
\begin{subequations}
	\begin{eqnarray}
		&\frac{\varepsilon^2q^2}{4}&=\bigg(\frac{\gamma}{(r^2-1)}+\frac{\varepsilon}{2}\bigg)^2+\bigg(\frac{\omega_0}{(r^2+1)}\bigg)^2,\\
		&\tan(2\psi)&=\frac{(1-r^2)(\omega_0)}{(r^2+1)(\gamma+\frac{\varepsilon}{2}(r^2-1))}.
	\end{eqnarray}
	\label{eq:stability-SSS}
\end{subequations}
$r$ and $\psi$ can be calculated  for a fixed set of parameters by numerically solving the above set of equations,
which is then 
substituted back into the evolution equations for the low-dimensional order parameters to deduce the characteristic equation.
The eigenvalues of the  characteristic equation  is then used to determine the saddle-node bifurcation curve in the suitable two parameter phase.

\subsection{Bimodal Distribution}
Now, we will deduce the low-dimensional evolution equations corresponding to the macroscopic order parameters  for the original
discrete set of equations~(\ref{eq:km2}) in the limit $N
\to \infty$  for the  asymmetric bimodal Lorentzian distribution function $g(\omega)$~(\ref{eq:bil}). Expanding the latter using partial fractions and
evaluating the integral in Eq.~(\ref{eq:13}) using appropriate contour integral, one can obtained the complex order parameter as
\begin{equation}
	z(t)=\frac{1}{2}[z_1(t)+z_2(t)],
\end{equation}
where
\begin{equation}
	z_{1,2}(t)=\alpha^\star(\pm\omega_0-i\gamma_{1,2},t).
\end{equation}

\begin{figure*}[!] 
		\centering
	\includegraphics[width=12cm]{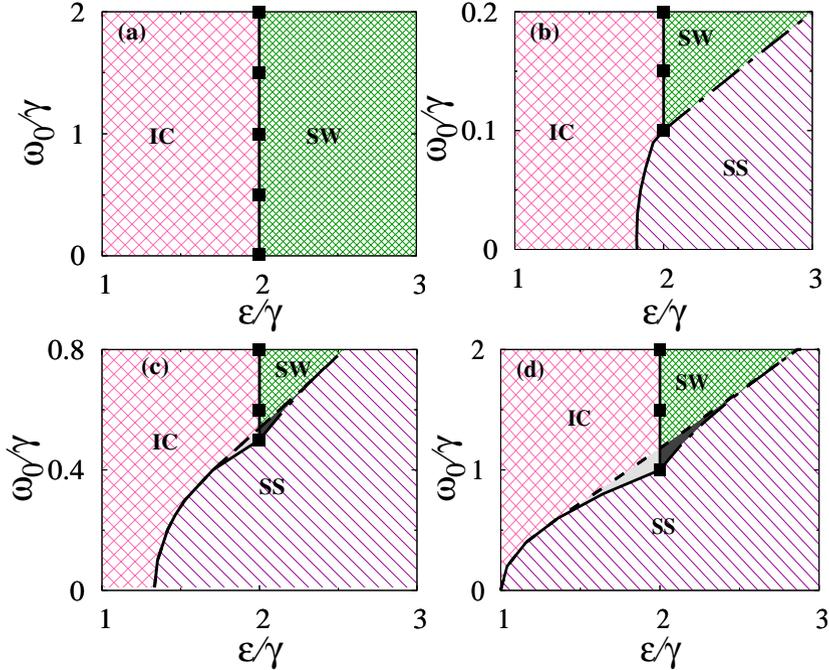}
	\caption{Phase diagrams in the $(\omega_0/\gamma, \varepsilon/\gamma)$ parameter space for the generalized Kuramoto  model (\ref{eq:km2}) with unimodal frequency distribution 
		for different values of the symmetry breaking parameter $q$.  (a) $q=0.0$, (b) $q=0.1$, (c) $q=0.5$, and (d) $q=1.0$.  The line connected by filled squares is
		the Hopf bifurcation curve $\varepsilon_{HB}$ (Eq.~(\ref{hb})), solid line corresponds to the pitchfork bifurcation curve  $\varepsilon_{PF}$  (Eq.~(\ref{eq:ISS}))
		dashed line corresponds to the saddle-node bifurcation curve (Eq.~(\ref{eq:stability-SSS})), and the dashed dotted line correspond to the homoclinic bifurcation curve
		obtained using the software XPPAUT. Bistability between the standing wave (SW) state and  the synchronized  stationary (SS) state is represented by dark shaded region enclosed by
		the saddle-node bifurcation  curve and the homoclinic bifurcation curve.  Bistability between the incoherent {(IC)} and the {SS}  state is represented by light grey shaded region enclosed by the saddle-node bifurcation  curve and  the pitchfork bifurcation curve.}
	\label{fig:i} 	
\end{figure*}
\begin{figure*}[!]	
	\hspace*{-1.0cm}
	\includegraphics[width=12.99cm]{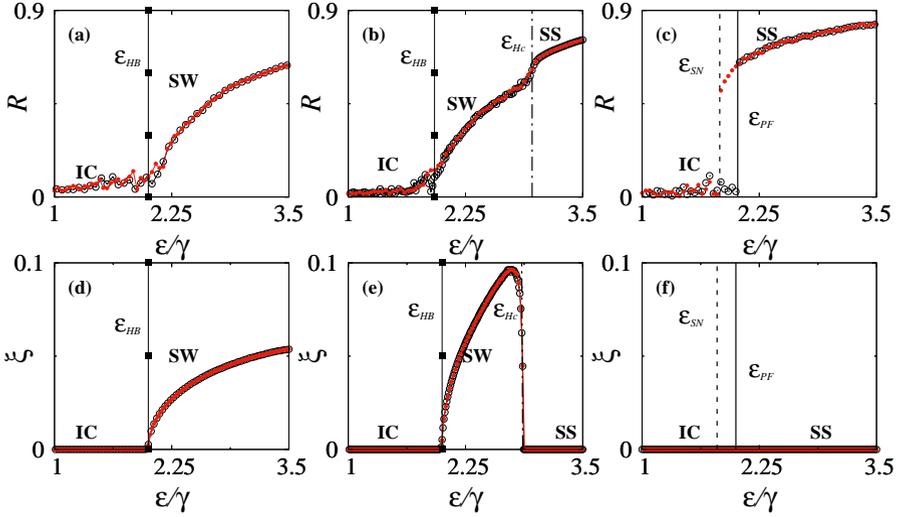}
	\caption{Time averaged order parameter $R$  and the Shinomoto-Kuramoto order parameter $\xi$ for the generalized Kuramoto model (\ref{eq:km2}) with unimodal frequency distribution as a function of  $\varepsilon/\gamma$
		for  $\omega_0/\gamma=1$. (a) and  (d) $q=0.0$ , (b)  and (e) $q=0.5$,  and (c)  and  (f) $q=1.0$.  The forward trace is indicated by the line connected by open circles, 
		while the reverse trace is indicated by the line connected by closed circles.   The states indicated by IC, SW and SS correspond to the incoherent, standing wave, 
		and synchronized stationary  states, respectively.  The bifurcation curves $\varepsilon_{HB}, \varepsilon_{Hc}, \varepsilon_{PF}$ and $\varepsilon_{SN}$ correspond to  the
		Hopf, homoclinic, pitchfork and saddle-node bifurcation curves, respectively.}
	\label{fig:ai}
\end{figure*}
Substitution it into Eq. (\ref{eq:12}) yields two
coupled complex ordinary differential equations describing the evolution of two suborder parameters as
\begin{align}
	\dot{z}_1=&-(\gamma_1+i\omega_0)z_1+\frac{\varepsilon}{4}[(z_1+z_2-(z_1^\star+z_2^\star)z_1^2)\nonumber\\&+q((z_1+z_2)z_1^2-(z_1^\star+z_2^\star))],\label{eq:z1}\\
	\dot{z}_2=&-(\gamma_2-i\omega_0)z_2+\frac{\varepsilon}{4}[(z_1+z_2-(z_1^\star+z_2^\star)z_2^2)\nonumber\\&+q((z_1+z_2)z_2^2-(z_1^\star+z_2^\star))].\label{eq:z2}
\end{align}
The above evolution equations  for the complex order parameters  $z(t)_{1,2}=r(t)_{1,2}e^{i\psi(t)_{1,2}}$ can be expressed in terms of  the evolution equations in $r_{1,2}$ and
$\psi_{1,2}$,  as 
\begin{subequations}
	\begin{align}
		\frac{{\rm d}r_1}{{\rm d}t}&=-\gamma_1
		r_1+\frac{\varepsilon}{4}\big[(1-r_1^2)(r_1+r_2\cos(\psi_2-\psi_1))\nonumber\\&+q((r_1^2-1)(r_1\cos(2\psi_1)+r_2\cos(\psi_2+\psi_1)))\big],~~~\\
		\frac{{\rm d}\psi_1}{{\rm
				d}t}&=-\omega_0+\frac{{\varepsilon}}{4r_1}(r_1^2+1)\big[r_2\sin(\psi_2-\psi_1)\nonumber\\&+q(r_1\sin(2\psi_1)+r_2\sin(\psi_2+\psi_1))\big].
	\end{align}
	\label{eq:r-bim}
\end{subequations}
and 
\begin{subequations}
	\begin{align}
		\frac{{\rm d}r_2}{{\rm d}t}&=-\gamma_2
		r_2+\frac{\varepsilon}{4}\big[(1-r_2^2)(r_1\cos(\psi_2-\psi_1)+r_2)\nonumber\\&+q((r_2^2-1)(r_1\cos(\psi_2+\psi_1)+r_2\cos(2\psi_2)))\big],~~~~\\
		\frac{d\psi_2}{dt}&=\omega_0-\frac{{\varepsilon}}{4r_2}(r_2^2+1)\big[r_1\sin(\psi_2-\psi_1)\nonumber\\&-q(r_1\sin(\psi_2+\psi_1)+r_2\sin(2\psi_2))\big].
	\end{align}
	\label{eq:r2-bim}
\end{subequations}
The above equations constitute the evolution equations for reduced low-dimensional  order parameters
corresponding to the dynamics~(\ref{eq:km2}) in the limit $N
\to \infty$ and for the case of the asymmetric bimodal Lorentzian  distribution $g(\omega)$~(\ref{eq:bil}). 
Now, we discuss the various  asymptotic macroscopic dynamical states
admitted by Eqs.~(\ref{eq:r-bim}) and (\ref{eq:r2-bim}). 


\subsubsection{Incoherent state}
The incoherent state is defined by $r_1$=$r_2$=0.
A linear stability analysis of the fixed point $(z_1,z_2)$ = (0, 0) results in the stability condition,
\begin{align}
	\omega_0^2=\frac{1}{4}(\varepsilon a_1-2a_2+\sqrt{\varepsilon^2q^2a_1-4\varepsilon a_3^2+4a_3^2a_1}),
	\label{eq:pf}
\end{align}
where, $a_1=\gamma_1+\gamma_2,  a_2=\gamma_1^2+\gamma_2^2$ and $a_3=\gamma_1-\gamma_2$.
This stability curve actually corresponds to the pitchfork bifurcation curve across which the fixed point $(z_1,z_2)$ = (0, 0) (incoherent state) loses its stability leading 
to the synchronized stationary state.  Note that the incoherent state  loses it stability  through  the Hopf bifurcation, which results in the stability condition
	\begin{align}
		\omega_0^2=&\frac{1}{4}(\varepsilon-2b_1)^4(\varepsilon^2(q^2-1)-16b_2+4\varepsilon b_1)^2\bigg[\varepsilon^5(q-1)b_1-\varepsilon^4(q^2-1)\big((q^2-8)b_3\nonumber\\&+2b_2(q^2-10)\big)-4\varepsilon^3(q^2-2)\big(3(\gamma_1^3+\gamma_2^3)+13b_2b_1\big)+4\varepsilon^2(b_1)^2\big(b_3(q^2-8)\nonumber\\&+2b_2(3q^2-20)\big)+16\varepsilon b_1^3(b_3+10b_2)-64b_2b_1^4\bigg],
		\label{eq:hb}
	\end{align}
where, $b_1=\gamma_1+\gamma_2, b_2=\gamma_1\gamma_2$ and  $b_3=\gamma_1^2+\gamma_2^2$.  The above stability curve
corresponds to the Hopf bifurcation curve.
The boundary of stable incoherent  state is  therefore  enclosed by both the pitchfork bifurcation and Hopf bifurcation curves.

\subsubsection{Synchronized stationary state}
Deducing the  solution for the synchronized stationary state for the  asymmetry bimodal distribution  may not be possible  as  $r_1~\ne~r_2$ and $\psi_1~\ne~\psi_2$. However, for the symmetry bimodal distribution characterized by $r_1~=~r_2$ and $\psi_1~=~-\psi_2$,  one can deduce the equations for $r$ and $\psi$ as in (\ref{eq:stability-SSS})
and  obtain the saddle-node bifurcation curves as pointed out in Sec.~\ref{subsec:sss}.

\begin{figure*}[!]
	\centering
	\hspace*{-1cm}
	\includegraphics[width=13cm,height=9.5cm]{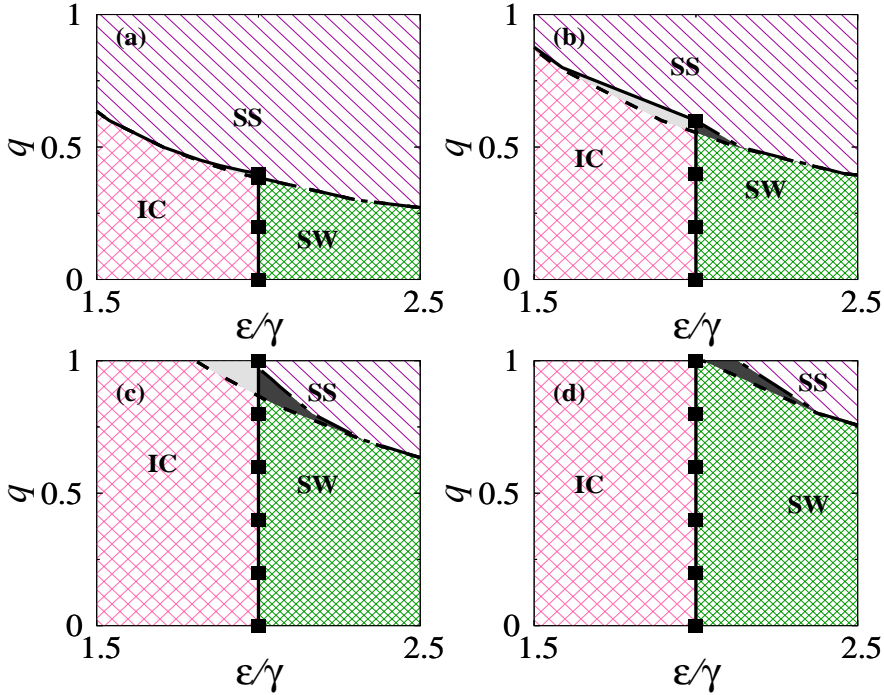}
	\caption{Phase diagrams in the $q-\varepsilon/\gamma$ parameter space for the generalized Kuramoto  model (\ref{eq:km2}) with unimodal frequency distribution 
		for increasing degree of  heterogeneity of the frequency distribution.  (a)  $\omega_0/\gamma_2=0.4$,  (b)  $\omega_0/\gamma_2=0.6$,  (c)  $\omega_0/\gamma_2=1.0$,  and
		(a)  $\omega_0/\gamma_2=1.2$.   The bifurcation curves and dynamical states are similar to those in Fig.~\ref{fig:i}.  }
	\label{fig:ii}
\end{figure*}
\begin{figure*}[!]
	\centering
	\hspace*{-1cm}
	\includegraphics[width=13cm]{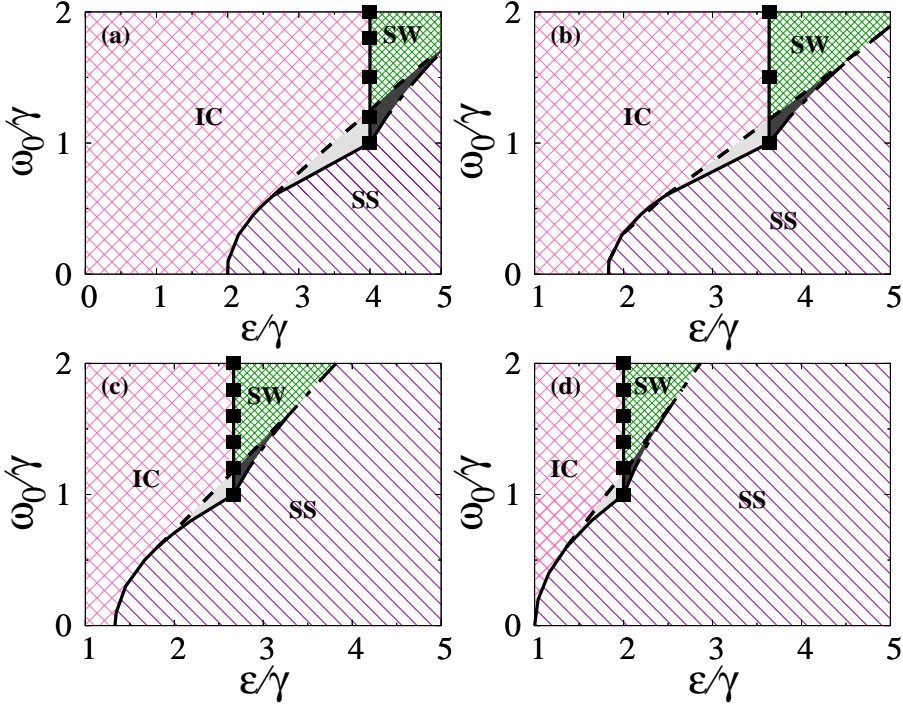}
	\caption{Phase diagrams in the $\omega_0/\gamma-\varepsilon/\gamma$ parameter space for the generalized Kuramoto  model (\ref{eq:km2}) with symmetric
		bimodal frequency distribution for increasing values of the strength of the symmetry breaking coupling. (a) $q=0.0$, (b) $q=0.5$, (c) $q=0.8$, and (d) $q=1.0$.  
		The bifurcation curves and dynamical states are similar to those in Fig.~\ref{fig:i}.}	
	\label{fig:iii}
\end{figure*}
\begin{figure}[!]
	\hspace*{-1cm}
	\centering
	\includegraphics[width=10cm]{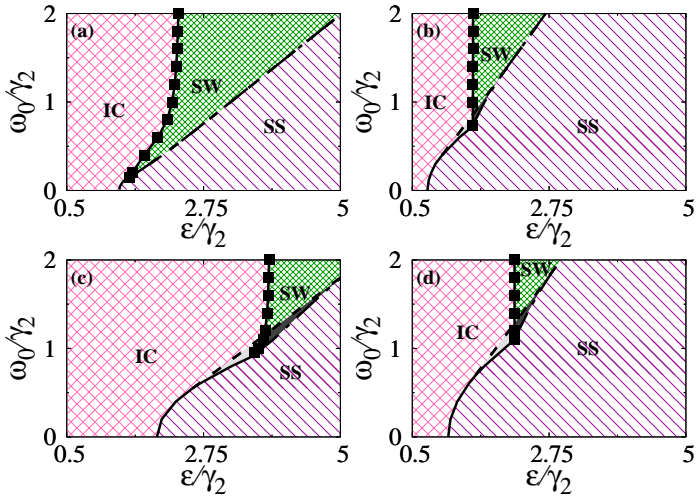}
	\caption{Phase diagrams in the $\omega_0/\gamma_2-\varepsilon/\gamma_2$ parameter space for the generalized Kuramoto  model (\ref{eq:km2}) with asymmetric
		bimodal frequency distribution for increasing  the strength of the symmetry breaking coupling and increasing the asymmetry between the bimodal frequency distributions. 
		(a)  and  (b) $\gamma_1/\gamma_2=0.6$,  and (c)  and  (d) $\gamma_1/\gamma_2=1.2$.  (a)  and  (c) $q=0.1$ and (b) and (d) $q=1$.  The bifurcation curves and 
		dynamical states are similar to those in Fig.~\ref{fig:i}. }	
	\label{fig:iv}	
\end{figure}
\begin{figure}[!]
	\hspace*{-1cm}
	\centering
	\includegraphics[width=10cm]{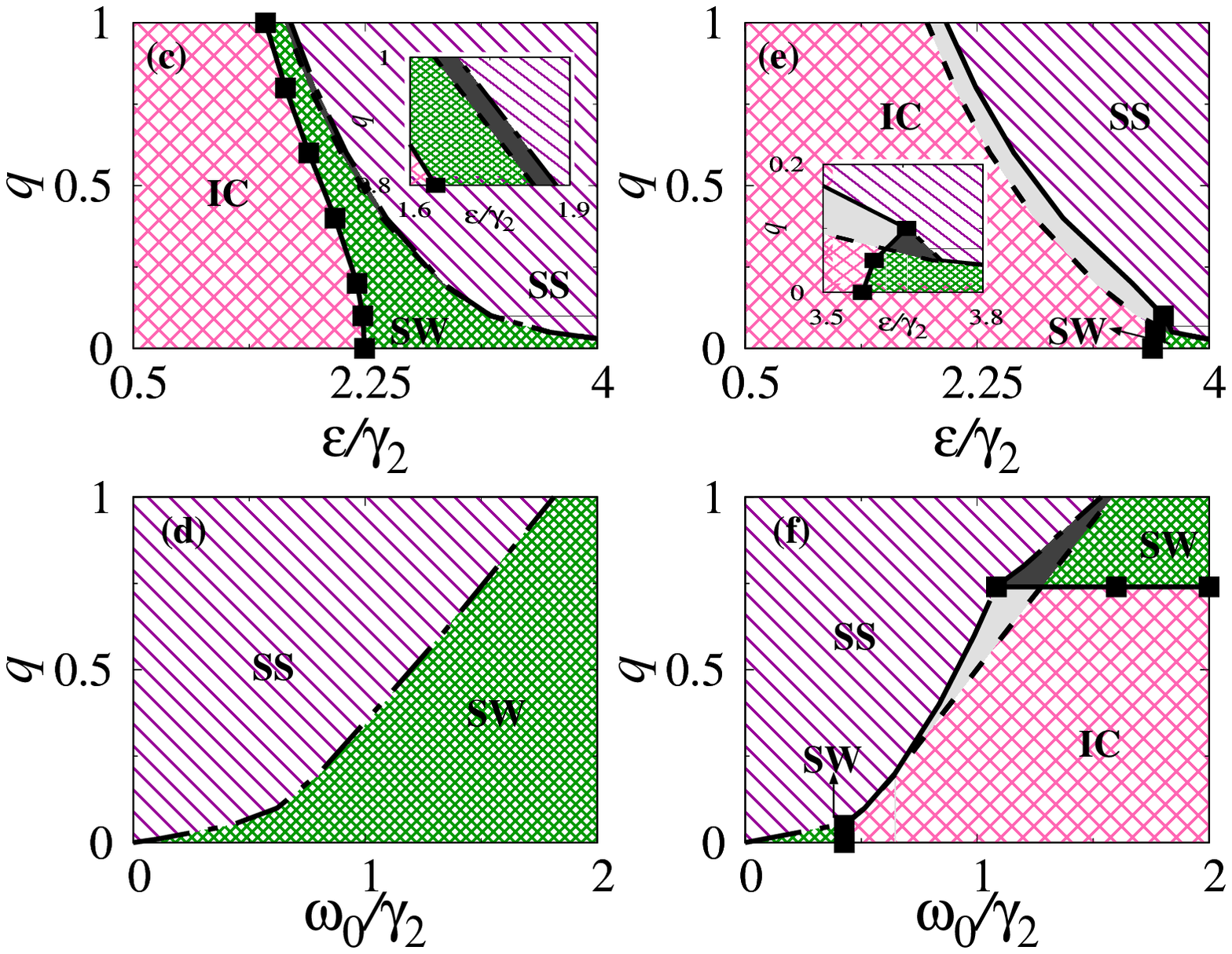}
	\caption{Phase diagrams in the $q-\varepsilon/\gamma_2$ parameter space (first row)  for $\omega_0/\gamma_2=1$ and  $q-\omega_0/\gamma_2$ (second row)  
		for $\varepsilon/\gamma_2=2.5$  for the generalized Kuramoto  model 
		(\ref{eq:km2}) with asymmetric bimodal frequency distribution.   (a)  and  (c)  $\gamma_1/\gamma_2= 0.6$,  and  (b)  and  (d)  $\gamma_1/\gamma_2= 1.2$.}
	\label{fig:v}	
\end{figure}

\section{Numerical Results}
In this section, we will proceed to unravel  the macroscopic dynamical states admitted by the generalized Kuramoto model (\ref{eq:km2}) with explicit symmetry breaking coupling
by constructing appropriate two parameter phase diagrams and classifying the underlying dynamical states from a numerical analysis of the
governing equations of the original  discrete model. Specifically, we will unravel the rich phase diagrams of the generalized Kuramoto model,
using both  unimodal and bimodal frequency distributions, for distinct values of the symmetry breaking parameter $q$. 
The number of oscillators is fixed as N = $10^4$, and we use the standard 4th-order Runge-Kutta integration scheme with integration step size h = 0.01
to solve the generalized Kuramoto model (\ref{eq:km2}).  
{Note that one can break the two-parameter phase into several segments  and multiple copies of the same code can be simulated simultaneously
	for different values of the parameters to generate the data, which can then be concatenated  to get the complete phase diagrams with a reasonable workstation.}

The initial state of the oscillators ($\theta_i$'s ) is distributed with uniform random values between -$\pi$ and +$\pi$. 
We use the time averaged order parameter $R$ defined as
\begin{equation}
	R=\lim_{t \to \infty}\frac{1}{\tau}\int_{t}^{t+\tau}r(t)dt,
	\label{eq:R}
\end{equation}
where $r(t)=\vert Z\vert=\vert N^{-1}\sum_{j=1}^{N}e^{i\theta_j} \vert$.
Incoherent state  is characterized by $R=r(t)=0$, while the synchronized stationary state  is  characterized by $R=r(t)=const.$ Standing wave is
characterized by the oscillating nature of $r(t)$.  In order to distinguish the synchronized stationary state and the standing wave state more clearly, we
use  the Shinomoto-Kuramoto order parameter~\cite{gallego,ssyk1986} 
\begin{align}
	\xi=\overline{\vert r(t) - R\vert},
\end{align}
where $\bar{z}$ denoted the long time average.   Shinomoto-Kuramoto order parameter takes $\xi=0$ for the incoherent and synchronized stationary states, 
whereas it takes nonzero value  for  the standing wave state.

\subsubsection{Phase diagrams for the unimodal distribution}
We have depicted phase diagrams in   the ($\omega_0/\gamma, \varepsilon/\gamma$) parameter space for different values of the symmetry breaking 
parameter $q$ in Fig.~\ref{fig:i} in order to understand the effect of the explicit symmetry breaking interaction on the dynamics of Eq. (\ref{eq:km2}) with
unimodal frequency distribution.  The phase diagram is demarcated  into different dynamical regions using the value of the time averaged order parameter $R$ and 
the Shinomoto-Kuramoto order parameter $\xi$.
Incoherent state (IC), synchronized stationary state (SS) and standing wave (SW), along with the bistable regions  (dark and light gray shaded regions) are observed  in the phase
diagram. The parameter space indicated by light gray shaded region corresponds to the bistable regime between the incoherent and the synchronized stationary states, 
while that  indicated  by dark gray shaded region corresponds  to the bistable regime between the standing wave and the synchronized stationary states, 

Only the incoherent and standing wave states are observed in the phase diagram for $q=0$ (see ~\ref{fig:i}(a)), a typical phase diagram of the Kuramoto model
with unimodal frequency distribution.  The line connected by the filled squares corresponds to the Hopf bifurcation curve, across which there is a transition
from the  incoherent state to  the standing wave  state. Note that a finite value of $q$ results in the loss of the  rotational symmetry of the dynamics of the Kuramoto oscillators.
Even a feeble value of $q$ manifests the synchronized stationary state in a rather large parameter space at the cost of the standing wave state (see Fig.~\ref{fig:i}(b) for q=0.1). 
There is a transition from the incoherent state to  the standing wave  state via
the Hopf bifurcation curve $\varepsilon_{HB}$ (indicated by the line connected by filled squares) as a function of $\varepsilon/\gamma$  for $\omega_0/\gamma>0.1$.   
The standing wave  state loses its stability via the homoclinic bifurcation (indicated by the dashed-dotted line) as a function of $\varepsilon/\gamma$ resulting 
in the synchronized stationary state.   There is also a transition from the incoherent state  to the synchronized stationary state for  $\omega_0/\gamma\le 0.1$ as a function of 
$\varepsilon/\gamma$ via the pitchfork bifurcation curve $\varepsilon_{PF}$  indicated by the solid line.  

Further larger values of the symmetry breaking parameter results in the emergence of the bistability between the standing wave and  the synchronized stationary states  (indicated by
dark shaded region) enclosed by the  saddle-node bifurcation curve (indicated by dashed line) and   the homoclinic bifurcation curve (see Fig.~\ref{fig:i}(c) for q=0.5).
There is also a bistable region between the  incoherent state and the  synchronized stationary state  (indicated by light grey shaded region) enclosed by the  
saddle-node bifurcation curve
and the pitchfork bifurcation curve.  For $q=1$, both the bistable regions enlarged in the phase diagram (see Fig.~\ref{fig:i}(d)), which is a typical phase diagram of
the Winfree model with the unimodal frequency distribution.  The phase diagrams for $q=0.5$ and $1.0$  have similar dynamics except for the regime shift and enhanced bistabilities
in a larger parameter space.
Thus, as the value of $q$ is increased from the null value to the unity, one can observe the transition from the phase diagram of  the Kuramoto model  to that of the Winfree model.  
Note that the Hopf, saddle-node and pitchfork bifurcation curves are the analytical bifurcation curves, Eqs.~(\ref{hb}), (\ref{eq:ISS}) and (\ref{eq:stability-SSS}) respectively,  obtained from the
low-dimensional evolution equations  for the order parameters deduced in  Sec.~\ref{sec:ud}.  Homoclinic bifurcation curve is obtained from the software XPPAUT~\cite{xpp}.

Time averaged order parameter $R$ and the Shinomoto-Kuramoto order parameter $\xi$  are depicted in Fig.~\ref{fig:ai} as a function of $\varepsilon/\gamma$  for different values of the symmetry breaking parameter $q$ and
$\omega_0/\gamma$.  The forward trace is indicated by the line connected by open circles, while the backward trace is indicated by the line
connected by closed circles. There is a smooth (second order) transition from the incoherent  to  the standing wave states  via the Hopf bifurcation  $\varepsilon_{HB}$
at  $\varepsilon/\gamma=2$  during  both forward and reverse traces for $q=0.0$ and  $\omega_0/\gamma=1$ as depicted in Figs.~\ref{fig:ai}(a) and \ref{fig:ai}(d).
In addition, to the 
smooth transition  from the incoherent state to  the standing wave  state  via the Hopf bifurcation $\varepsilon_{HB}$ at  $\varepsilon/\gamma=2$,  there is  another 
smooth transition from the standing wave  state to the synchronized stationary state via the homoclinic bifurcation $\varepsilon_{Hc}$ at $\varepsilon/\gamma=2.94$ in both the 
forward and reverse traces  as shown in Fig.~\ref{fig:ai}(b) for $q=0.5$ and $\omega_0/\gamma=1$.  The transition from  the standing wave state to the 
synchronized stationary state is also corroborated by the sharp fall of the Shinomoto-Kuramoto order parameter $\xi$ to the null value (see Fig.~\ref{fig:ai}(e)).
In contrast, there is an abrupt (first order) transition 
from the  incoherent state to the synchronized stationary state at  $\varepsilon/\gamma=2$ via the pitchfork bifurcation curve $\varepsilon_{PF}$  
for $\omega_0/\gamma=1$ during the forward trace, whereas there is an abrupt transition from the synchronized stationary state to the incoherent state at 
$\varepsilon/\gamma=1.8$ via the saddle-node bifurcation  $\varepsilon_{SN}$ during the reverse trace (see  Fig.~\ref{fig:ai}(c) for $q=1.0$)
elucidating the presence of hysteresis and bistability between  the  incoherent state and  the synchronized stationary state.  The Shinomoto-Kuramoto order parameter $\xi$ 
takes the null value, in the entire range of $\varepsilon/\gamma$ in  Fig.~\ref{fig:ai}(f) for $q=1.0$,  characterizing both the incoherent and  the synchronized stationary states.

The observed dynamical states and their transitions are depicted in the $(q, \varepsilon/\gamma)$ parameter space for different  $\omega_0/\gamma$ in Fig.~\ref{fig:ii}.
The bifurcations mediating the dynamical transitions are similar to those observed in  Fig.~\ref{fig:i}.   The phase diagram for $\omega_0/\gamma=0.4$ is shown 
in Fig.~\ref{fig:ii}(a).  There is a transition from the  incoherent state  to  the standing wave state  via the Hopf bifurcation curve for smaller values of
the symmetry breaking parameter as a function of $\varepsilon/\gamma$. Larger values of the symmetry breaking parameter   favor  the 
synchronized stationary state in the entire range of  $\varepsilon/\gamma$.  However, in a narrow range  of  $q\in(0.36, 0.46]$ (see  Fig.~\ref{fig:ii}(a)),
there is a transition from  the incoherent state  to  the standing wave state and then to  the synchronized stationary state.  There is also a transition from 
the incoherent state to the synchronized stationary state in the range of  $q\in(0.46, 0.6)$. Recall that $\omega_0$  quantifies the degree of detuning of the frequency distribution.  
Increase in the heterogeneity of the frequency distribution promotes bistable regions,  incoherent and standing wave states,
to a large region of the  $(q, \varepsilon/\gamma)$ parameter space.  For instance, the phase diagram for  $\omega_0/\gamma=0.6$ is  depicted in Fig.~\ref{fig:ii}(b)
elucidates the emergence of the bistable regions and enlarged regions of the incoherent and standing wave states as a function of $q$, 
a manifestation of increased heterogeneity. Further increase in the  $\omega_0/\gamma$ enlarges the bistable regions,  the incoherent and the standing wave  
states as depicted in Figs.~\ref{fig:ii}(c) and ~\ref{fig:ii}(d) for
$\omega_0/\gamma=1$ and $1.2$, respectively.  These results are in agreement with the phase diagrams  in  Fig.~\ref{fig:i} in  the  $(\omega_0/\gamma, \varepsilon/\gamma)$
parameter space for increasing values of the  symmetry breaking parameter.  Next, we will explore the effect of symmetric and asymmetric bimodal frequency
distributions on the phase diagrams in the following.

\subsubsection{Phase diagrams for bimodal distribution}

In this section, we analyse the phase space dynamics of  the generalized Kuramoto model (\ref{eq:km2}) with symmetric bimodal frequency distribution~(\ref{eq:bil}) by setting
$\gamma=\gamma_1=\gamma_2$  for 
increasing values of the strength of the symmetry breaking coupling. We have depicted the phase diagrams in the $(\omega_0/\gamma, \varepsilon/\gamma)$ parameter space 
for different values of the symmetry breaking  parameter $q$ in Fig.~\ref{fig:iii}. Note that the phase space dynamics of
the Kuramoto model (see Fig. ~\ref{fig:iii}(a) for $q=0$)  are similar to those of the Winfree model  (see Fig. ~\ref{fig:iii}(d) for $q=1$)   for the symmetric bimodal  frequency distribution
except for the regime shift. The dynamical states and the bifurcation curves are similar to those in Fig.~\ref{fig:i}.  Increasing the strength of the  symmetry breaking coupling
favors the  synchronized stationary state and the bistable states in a  large region of the parameter space as evident from  Fig. ~\ref{fig:iii}(b) and  ~\ref{fig:iii}(c)  for  
$q=0.5$ and $q=0.8$, respectively. Note that  a large heterogeneity in the frequency distribution favor  the incoherent  and the  standing wave states in a rather large region 
of the phase diagram for smaller $q$  and $\varepsilon$ (see Fig. ~\ref{fig:iii}(a) for $q=0$). Nevertheless, the  synchronized stationary state  predominates the 
phase diagram for larger strength of the symmetry breaking coupling and $\varepsilon$ despite the presence of a large heterogeneity in the frequency distribution 
(see Fig. ~\ref{fig:iii}(d) for $q=1$). 

Next, we analyze the phase space dynamics of the generalized Kuramoto model (\ref{eq:km2}) with asymmetric bimodal frequency distribution~(\ref{eq:bil}) by
increasing  the strength of the symmetry breaking coupling and  the  degree of asymmetry between the bimodal frequency distributions.
We have depicted the phase diagrams in the $(\omega_0/\gamma_2, \varepsilon/\gamma_2)$ parameter space 
for different values of the symmetry breaking  parameter $q$ in Fig.~\ref{fig:iv}.  
Again, the dynamical states and the bifurcation curves are similar to those in Fig.~\ref{fig:i}.   Phase diagram for $q=0.1$ and $\gamma_1/\gamma_2=0.6$ 
is depicted in Fig.~\ref{fig:iv}(a).   For most values of $\omega_0/\gamma_2$, there is a transition from the incoherent state to the  synchronized stationary state via 
the standing wave state and there is no bistability
for $\gamma_1<\gamma_2$.  However,  there is a transition from the incoherent state to the  synchronized stationary state  in  a large range of  $\omega_0/\gamma\in(0,1)$
and the emergence of bistable states for $\gamma_1>\gamma_2$ as depicted in Fig.~\ref{fig:iv}(b) for   $\gamma_1/\gamma_2=1.2$. It is evident that
bistable states emerge even for  low values of the symmetry breaking coupling  when $\gamma_1>\gamma_2$.  
Note that bistable states emerge even for  $\gamma_1<\gamma_2$ but for a large strength of the symmetry breaking coupling  (see Fig.~\ref{fig:iv}(c)
for $q=1$ and $\gamma_1/\gamma_2=0.6$).  The spread of the bistable states increases for $q=1$ and  $\gamma_1/\gamma_2=1.2$ as illustrated in 
Fig.~\ref{fig:iv}(d).  Thus, larger  $\gamma_1/\gamma_2$ and $q$ favor the  emergence of the bistable states.

Phase diagrams in the  $(q, \varepsilon/\gamma_2)$  parameter space is depicted in Figs.~\ref{fig:v}(a) and ~\ref{fig:v}(b) for  $\gamma_1/\gamma_2=0.6$  and $1.2$, respectively,
and for $\omega_0/\gamma_2=1$.  The dynamical states and the bifurcation curves are similar to those in Fig.~\ref{fig:i}.   
There is a transition from the   incoherent state to the  synchronized stationary state  via  the standing wave state for small values of $q$  (see Fig..~\ref{fig:v}(a)) 
similar to that in Fig.~\ref{fig:iv}(a).  However, for larger values of $q$
multistability between the standing wave  and  the  synchronized stationary state   emerges (dark shaded region in the inset)  in addition to the above dynamical transition.   
For  $\gamma_1>\gamma_2$, there  a transition from the incoherent state  to the standing wave state along with the  bistability among them in 
a rather narrow range of $q\in (0,0.4)$ as a function of
$\varepsilon/\gamma_2$ as shown in inset of Fig.~\ref{fig:v}(b).  For $q> 0.4$, there is a transition from the incoherent state  to the synchronized stationary state 
with the onset of bistability (light grey shaded region)
between them. Phase diagrams in the  $(q, \omega_0/\gamma_2)$  parameter space is depicted in Figs.~\ref{fig:v}(c) and ~\ref{fig:v}(d) for 
$\gamma_1/\gamma_2=0.6$  and $1.2$, respectively, for $\varepsilon/\gamma_2=2.5$.  There is a transition from the  synchronized stationary state 
to the  standing wave state as a function  of $\omega_0/\gamma_2$
for  $\gamma_1<\gamma_2$ (see Fig.~\ref{fig:v}(c)) via the homoclinic bifurcation curve.   Both the bistable states emerge when $\gamma_1>\gamma_2$  
as shown in Fig.~\ref{fig:v}(c) for $\gamma_1=1.2$.\\

\section{Conclusions}
\label{sec:conclusions}
We have considered a nontrivial generalization of the paradigmatic Kuramoto model by using an additional coupling term that explicitly breaks the rotational symmetry
of the Kuramoto model.   The strength of the symmetry breaking coupling is found to play a key role in the manifestation of  the dynamical states and their transitions 
along with the onset of bistability among the observed dynamical states in the phase diagram.
A typical phase diagram of the Kuramoto model  is transformed into a typical  phase diagram  of the Winfree mode
for the unit value of the  strength of the symmetry breaking coupling  thereby bridging the dynamics of both the Kuramoto and Winfree models.
Large values of the strength of the symmetry breaking coupling favor the manifestation of bistable regions and synchronized stationary state in a large region 
of the phase diagram. The dynamical transitions in the bistable region are characterized by an abrupt (first-order) transition in both the forward and reverse traces. 
Phase diagrams of both the Kuramoto and Winfree models resemble each other for symmetric bimodal frequency distribution except for the regime shifts and the degree of
the spread of the dynamical states and bistable regions.  Nevertheless, for asymmetric bimodal frequency distribution one cannot observe the bistable states
for  low values of the strength of the symmetry breaking coupling when  $\gamma_1<\gamma_2$.   In contrast, 
bistable states emerge even for  $\gamma_1<\gamma_2$ for a large strength of the symmetry breaking coupling. 
Larger  $\gamma_1/\gamma_2$ and  larger $q$ favors the  emergence of the bistable states in the case of the asymmetric  bimodal frequency distribution.
A large  $\omega_0$ and consequently a large degree of heterogeneity  facilitates the  spread of the incoherent and standing wave states  in the phase diagram for a low strength of
the  symmetry breaking coupling. However, a large $q$ promotes the spread of the  synchronized stationary state and bistable regions in the phase diagram 
despite the degree of heterogeneity in the frequency distribution.  We have deduced the low-dimensional evolution equations for the complex order parameters
using the  Ott-Antonsen ansatz for both unimodal and bimodal frequency distributions.  
We have also deduced the Hopf, pitchfork, saddle-node bifurcation curves  from the low-dimensional evolution equations for the complex order parameters.
 Homoclinic bifurcation curve is obtained from XPPAUT software. Simulation results, obtained from the original discrete set of  
equations agrees well with the analytical bifurcation curves. We sincerely believe that our results  will 
shed more light and enhance  our current understanding of the effects of symmetry breaking coupling in the phase models  and bridges the dynamics of
two distinctly different phase models,  which are far from reach otherwise.\\

\section{Acknowledgements}

The work of V.K.C. is supported by the DST-CRG Project under Grant No. CRG/2020/004353 and  
DST, New Delhi for computational facilities under the  DST-FIST program (SR/FST/PS- 1/2020/135)to the 
Department of Physics. M.M. thanks the Department of
Science and Technology, Government of India, for provid-
ing financial support through an INSPIRE Fellowship No.
DST/INSPIRE Fellowship/2019/IF190871.
S.G. acknowledges support from the Science
and Engineering Research Board (SERB), India under SERB-TARE scheme Grant No.
TAR/2018/000023 and SERB-MATRICS scheme Grant No. MTR/2019/000560. He also thanks ICTP -- The Abdus Salam International Centre for Theoretical Physics,
Trieste, Italy for support under its Regular Associateship scheme.   DVS  is supported by the DST-SERB-CRG Project under Grant No. CRG/2021/000816.\\

\textbf{Data Availability Statement}: No Data associated in the manuscript. The data sets on the current study are available from the corresponding author on reasonable request.


\end{document}